\documentclass[aip,rsi,reprint,graphicx]{revtex4-1} 
\usepackage{citesort}

\usepackage{graphicx}
\draft 

\begin{document}


\title{A table-top, repetitive pulsed magnet for nonlinear and ultrafast spectroscopy in high magnetic fields up to 30~T} 



\author{G. Timothy Noe II}
\affiliation{Department of Electrical and Computer Engineering, Rice University, Houston, Texas 77005, USA}

\author{Hiroyuki Nojiri}
\affiliation{Institute for Materials Research, Tohoku University, Sendai 980-8577, Japan}

\author{Joseph Lee}
\affiliation{Department of Electrical and Computer Engineering, Rice University, Houston, Texas 77005, USA}

\author{Gary L. Woods}
\affiliation{Department of Electrical and Computer Engineering, Rice University, Houston, Texas 77005, USA}

\author{Jean L\'{e}otin}
\affiliation{Laboratoire National des Champs Magn\'{e}tiques Intenses, CNRS-UJF-UPS-INSA, Toulouse,
France}

\author{Junichiro Kono}
\email[]{kono@rice.edu}
\homepage[]{http://www.ece.rice.edu/~kono}
\affiliation{Department of Electrical and Computer Engineering, Rice University, Houston, Texas 77005, USA}
\affiliation{Department of Physics and Astronomy, Rice University, Houston, Texas 77005, USA}
\affiliation{Department of Materials Science and NanoEngineering, Rice University, Houston, Texas 77005, USA}


\date{\today}

\begin{abstract}
We have developed a mini-coil pulsed magnet system with direct optical access, ideally suited for nonlinear and ultrafast spectroscopy studies of materials in high magnetic fields up to 30~T.  The apparatus consists of a small coil in a liquid nitrogen cryostat coupled with a helium flow cryostat to provide sample temperatures down to below 10~K.  Direct optical access to the sample is achieved with the use of easily interchangeable windows separated by a short distance of $\sim$135~mm on either side of the coupled cryostats with numerical apertures of 0.20 and 0.03 for measurements employing the Faraday geometry.  As a demonstration, we performed time-resolved and time-integrated photoluminescence measurements as well as transmission measurements on InGaAs quantum wells.
\end{abstract}


\maketitle 

\section{Introduction}
\label{sec:Intro}
Combining access to applied magnetic fields with ultrafast spectroscopy techniques and/or intense, pulsed laser sources can provide a wealth of information in condensed matter physics, including many-body interactions and Coulomb correlations in semiconductors in the quantum Hall regime,\cite{KneretAl98PRL,FromeretAl02PRL,DanietAl06PRL} optical properties of exotic materials and/or semiconductor magneto-plasmas in the terahertz frequency range,\cite{KonoetAl99APL,ZudovetAl03JAP,KhodaparastetAl03PRB,WangetAl07OL,IkebeetAl10PRL,WangetAl10NP,WangetAl10OE,TakahashietAl11NP,ArikawaetAl11PRB,BordacsetAl12NP,ArikawaetAl12OE,ShimanoetAl13NC} and ultrafast spectroscopy and control of magnetization in magnetic semiconductors and ferromagnets.\cite{BeaurepaireetAl96PRL,KojimaetAl03PRB,WangetAl06JPC}

Typically, experimental access to magnetic fields up to 30~T are limited to special facilities in the form of national laboratories.\cite{HerlachMiura03Book,CrowetAl96Physica,SingletonetAl04Physica,KiyoshietAl06JPCS,WosnitzaetAl07JMMM,LietAl10JLTP,DebrayFrings13CRP}.  Specifically, picosecond and femtosecond spectroscopy experiments in high magnetic fields have been limited to special magnets and/or facilities within high magnetic field laboratories,\cite{HeberleetAl95Physica,BhowmicketAl12PRB} including the Ultrafast Optics Facility at the National High Magnetic Field Laboratory.  This facility has recently developed a new direct current (DC) high field magnet, the Split-Florida Helix,\cite{TothetAl12IEEE} which combines direct optical access via four elliptical window ports with magnetic fields up to 25~T to greatly expand the number of possible spectroscopic experiments utilizing high magnetic fields.  However, the somewhat large size of the magnet, $\sim$1~meter in diameter, makes it difficult to couple to optical experiments and forces the magnet to be the centerpiece of the experiment around which all the other optics are built.  Also, this new magnet is part of a user facility, making access limited to short periods of time during a user visit.  The major advantage of the Split-Florida Helix or other DC magnets over pulsed magnets is the fact that it provides a constant magnetic field, allowing for rapid signal averaging in sensitive measurements.  Elsewhere, methods have been developed to perform ultrafast terahertz measurements in high magnetic fields.  However, optical access is often limited to the use of optical fibers to couple into and out of the magnet system.\cite{Crooker02RSI,MolteretAl10OE}

Here, we have developed a mini-coil pulsed magnet\cite{NarumietAl12SRN} system that couples low temperatures, high magnetic fields, and direct optical access for use in a university lab setting.  The mini-coil design allows one to incorporate the magnet into a spectroscopy setup by placing it directly on the table-top. We present time-integrated and time-resolved photoluminescence (PL) results upon intense excitation with an amplified Ti:sapphire laser as well as absorption and weak excitation PL results using a light emitting diode (LED) and a laser diode, respectively, to demonstrate the utility of the mini-coil pulsed magnet for high-field magneto-optical spectroscopy.

\section{Experimental Setup}
\label{sec:ExpSetup}

\subsection{Mini-coil magnet system}
\label{sec:MagSys}

\begin{figure*}
\includegraphics[scale=0.86]{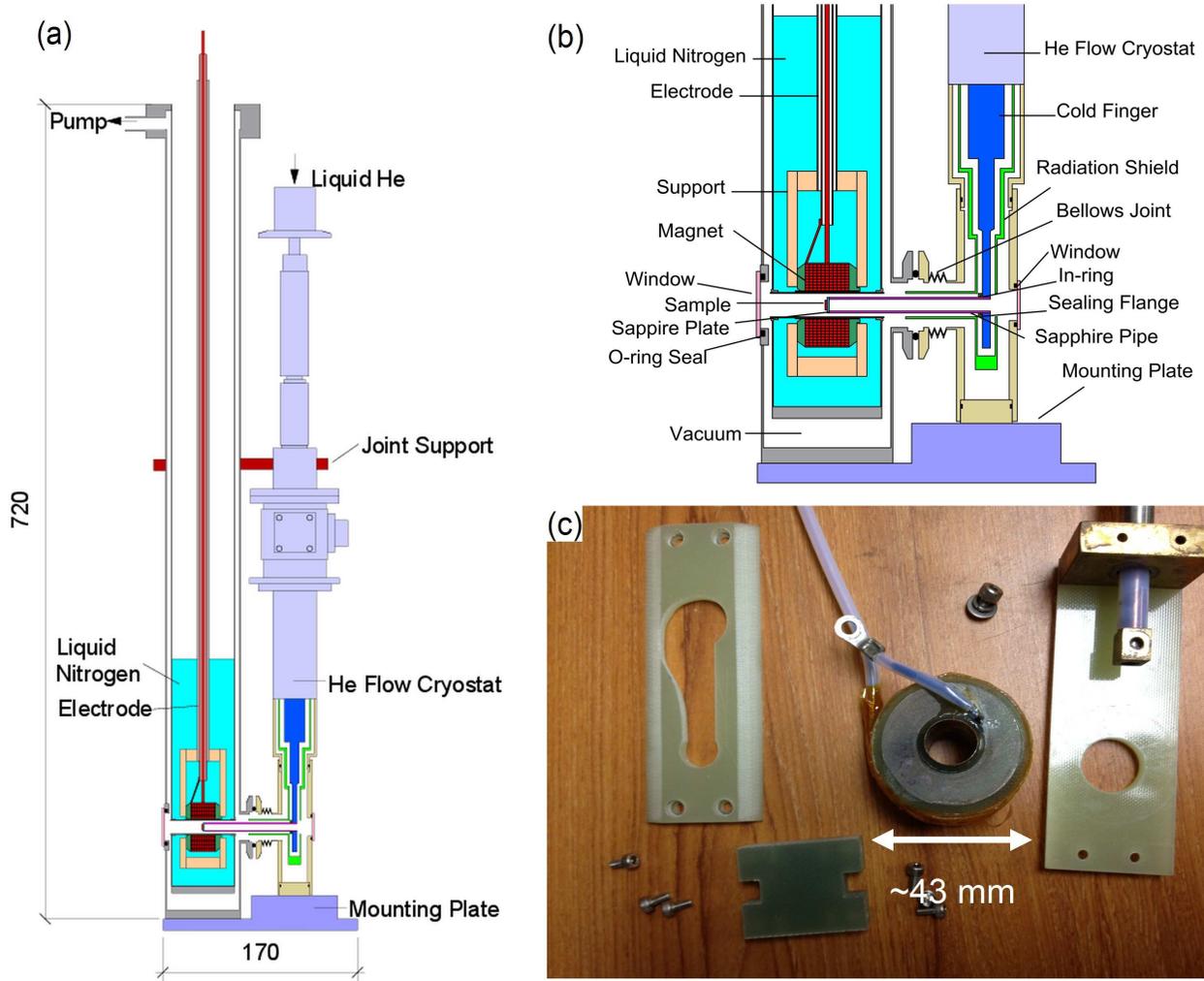}
\caption{Schematic diagram demonstrating the coupled cryostats and the various components of the magnet system.  (a)~The two cryostats, connected together on a mounting plate, 170~mm by 65~mm, that can be secured directly onto an optical table. (b)~An enlarged view of (a) showing the sample in the center of the magnet coil sitting at the end of a cylindrical sapphire pipe, which is secured to the copper cold finger of the helium flow cryostat using an indium ring and a sealing flange.  The sample can be mounted on either side of the sapphire plate.  Windows provide direct optical access on either side of the system.  (c)~A photo showing the magnet with an outer diameter of $\sim$43~mm.  A coaxial electrode connected to the mini-coil to deliver up to $\sim$5~kA to the coil generating a peak magnetic field of 30~T.}
\label{ill:Cryostat}
\end{figure*}

Our magnet system consists of two coupled cryostats, as shown in Fig.~\ref{ill:Cryostat}(a).  One cryostat contains a small magnet coil that must be kept in a bath of liquid nitrogen to cool the coil after each magnet shot.  The other cryostat is a commercial liquid helium flow cryostat (Cryo Industries, Inc., CFM-1738-102), which is used to cool the sample.  A cylindrical sapphire pipe extends from the helium flow cryostat's cold finger into the magnet bore to locate the sample in the peak magnetic field and cool the sample to around 10~K.  The sapphire pipe is held in place by clamping indium wire around the pipe with a two-piece copper sealing flange, which has a wedge cut on one of the pieces so that, when combining the two pieces of the flange, the indium is squeezed between the copper and sapphire to provide a secure fit.  The flange screws into the helium flow cryostat's cold finger.  One factor on which the minimum sample temperature achieved depends is the thermal connection between the copper, indium, and sapphire.  The lowest sample temperature that we have achieved so far is 8.4~K. The two cryostats share a common vacuum space and fit together on a mounting plate that is secured directly onto an optical table.  A pickup coil is wrapped around the sapphire pipe close to the sample location to accurately measure the generated magnetic field.  A temperature sensor is also located at the end of the sapphire pipe on its outer diameter, less than 5~mm from the sample.  Another temperature sensor is located at the end of the helium flow cryostat's copper cold finger, which can achieve a minimum temperature of 7~K.

Direct optical access is achieved by using 1-inch-diameter windows on either side of the magnet system.  These windows can be chosen for experiments, utilizing a variety of wavelengths.  The magnet bore diameter is 12~mm which limits the numerical aperture, $NA=\sin{\theta}$, to 0.20 on the magnet side with respect to the sample position, where $\theta=11.7^\circ$ is the half-angle of a cone of light from a point source at the sample position to the window.  On the helium flow cryostat side, the sapphire pipe limits the $NA$ to 0.03 as the pipe has an inner diameter of 6~mm and is 90~mm long, and the half-angle, $\theta=1.9^\circ$.  For our experiments, the laser beam enters the cryostat through a window on the helium flow cryostat side.  For comparison, the elliptical window ports on the Split-Florida Helix subtend a half-angle of 5.7$^\circ$ vertically and 22.5$^\circ$ horizontally.\cite{TothetAl12IEEE}

The magnet coil is made from rectangular wire, with a cross-sectional area of 1$\times$1.5~mm$^2$, made of a 50\%-50\% alloy of silver and copper. The coil contains 14-15 turns per layer on average and 13 layers, making a length of $\sim$22.5~mm with inner diameter of 14~mm and outer diameter of 43~mm; the inductance, $L$, of the coil is 408~$\mu$H.  It is connected to a capacitor bank with capacitance $C$ = 5.6~mF, which can store and deliver over 9~kJ of energy into the coil to generate a maximum peak magnetic field of 30~T using a peak current of almost 5.1~kA when the bank is charged up to $\sim$1.8~kV.  The capacitor bank and power supply with dimensions 0.7~m by 0.45~m by 1.4~m sits on rolling casters, making the unit portable.  

\begin{figure}
\includegraphics[scale=0.74]{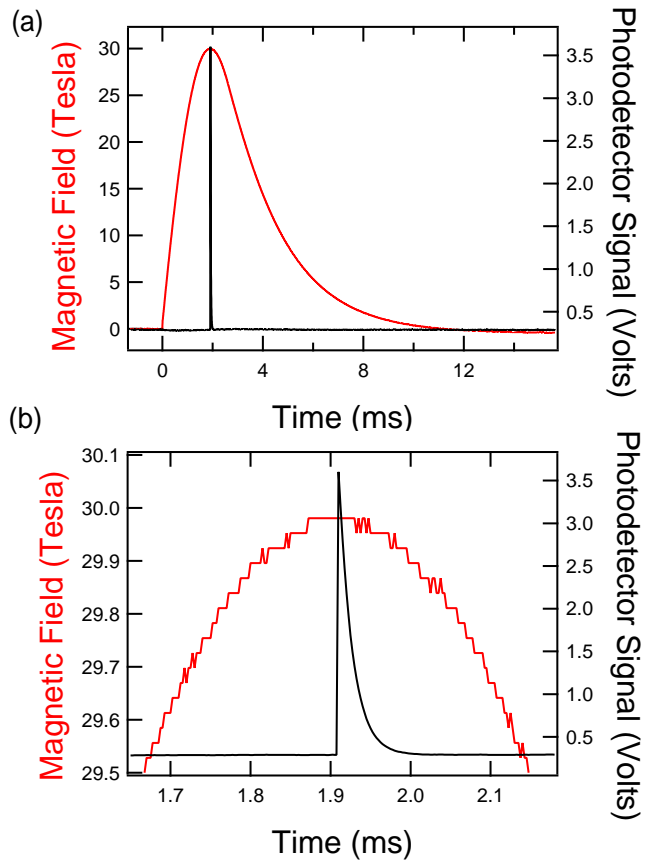}
\caption{(a)~Magnetic field profile for a 30~T shot (red) and photodetector signal measuring the laser excitation pulse (black).  A single excitation pulse arrives at the peak of the magnetic field.  (b)~The magnetic field variation at the top of the pulse is $\sim$4 parts in 300 for 400~$\mu$s and negligible for less than 1~$\mu$s.}
\label{ill:Sync}
\end{figure}

The magnetic field pulse profile resembles a half sine wave with exponential decay with a full width at half maximum of 4-5~ms (see Fig.~\ref{ill:Sync}), which is close to $\pi\sqrt{LC}$.  After discharging the stored energy of the capacitor bank into the coil, we must wait minutes, depending on the peak magnetic field strength, for the coil to cool back down to liquid nitrogen temperature before firing another magnet shot.  We can then repeat the measurement and average the data.  Relative to larger pulsed magnet coils, this mini-coil magnet can operate at a higher repetition rate.  For a 30~T shot, we have conservatively estimated a safe wait time of 10~minutes.  The repetition rate at which we can make magnet shots increases with decreasing peak magnetic field/current.  A control unit with synchronization and interlock functionality is used to operate the main magnet circuit. With this unit, we can specify the capacitor bank charge voltage, delay the time of the magnet pulse relative to the timing of other equipment, and prevent a magnet pulse if the interlock system is tripped.

\subsection{Optical spectroscopy methods}
\label{sec:OptSet}

\subsubsection{Time-resolved photoluminescence spectroscopy}

\begin{figure*}
\includegraphics[scale=0.84]{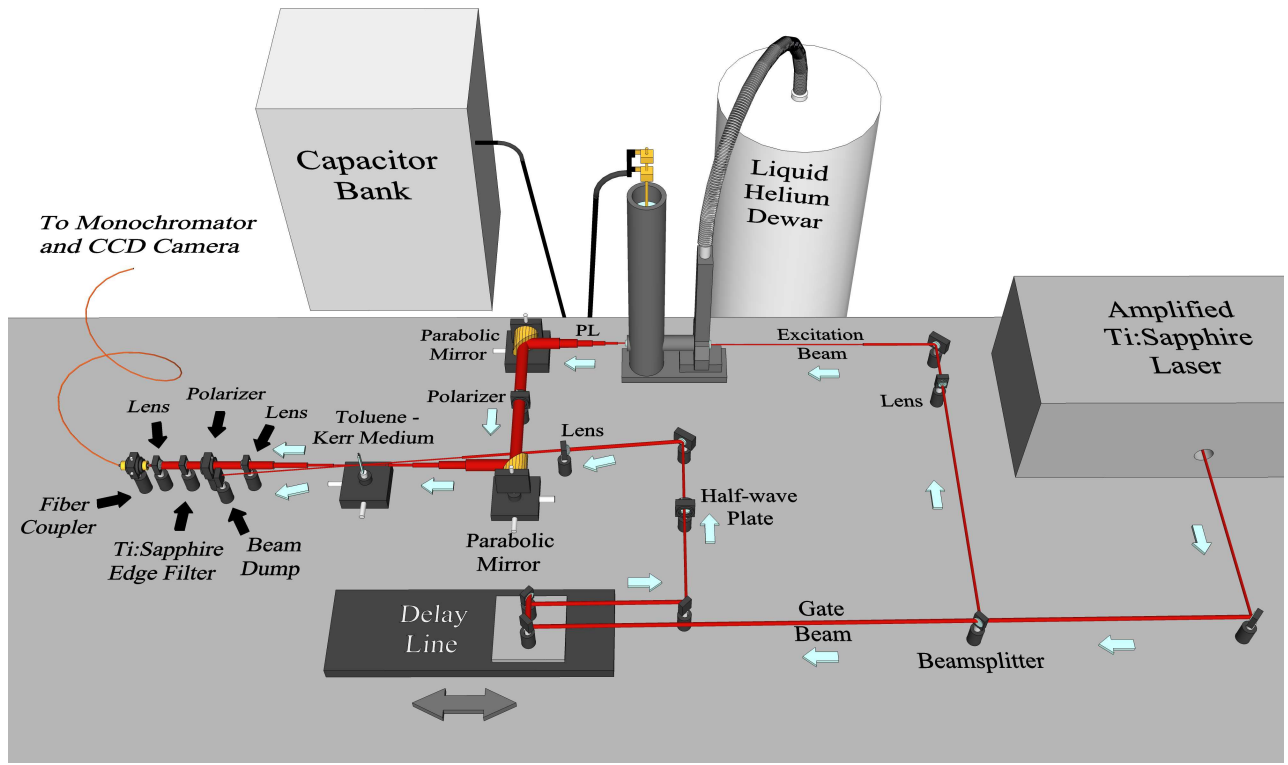}
\caption{Illustration of the optical Kerr gate setup for time-resolved photoluminescence experiments.  Light blue arrows indicate the direction of the light.  After excitation through a window on the helium flow cryostat side with the output of an amplified Ti:sapphire laser, the photoluminescence is collected through the window on the magnet cryostat side and focused onto a Kerr medium located between two crossed polarizers where it overlaps in space with a gate pulse.  The gate pulse causes transient birefringence in the medium so the polarization of the photoluminescence that passes the medium during the medium's response to the gate is rotated and can then partially pass through the second polarizer.  After making a series of magnet shots with an incrementally changing time delay between the excitation and gate pulses, the time-resolved photoluminescence intensity can be mapped as a function of time and wavelength.}
\label{ill:OpticalLayout}
\end{figure*}

We have incorporated the mini-coil magnet system into an ultrafast spectroscopy setup, as schematically shown in Fig.~\ref{ill:OpticalLayout}. 
In order to excite the sample, we use the output of the amplified Ti:sapphire laser (Clark-MXR, Inc., CPA-2001) centered at 775~nm with 1~kHz repetition rate, 150~fs pulse-width, and pulse energies up to 5~$\mu$J.  The excitation beam enters through the window on the helium flow cryostat side to optically excite the sample.  The sample sits on a sapphire plate on the helium flow cryostat side (shown opposite in Fig.~\ref{ill:Cryostat} for clarity) for the measurements demonstrated in this work.  The spot size of the excitation beam on the sample is $\sim$500~$\mu$m.  We use an optical chopper to reduce the repetition rate of the excitation pulses to 50~Hz because the silicon charge-coupled device (CCD) used to measure the emitted light cannot operate as fast as 1~kHz.  

We measure a split-off portion of the excitation laser pulses by using a silicon photodiode and use the leading edge of the 50~Hz signal as the main trigger source for the magnet's control unit.  We delay the beginning of the magnetic field generation/discharge of the capacitor bank by $\sim$18~ms after the one laser pulse to synchronize the excitation pulse with the peak of the magnetic field (see Fig.~\ref{ill:Sync}); the next excitation pulse arrives at the peak of the magnetic field.  This method allows us to measure the PL generated from a single excitation pulse at the peak of the magnetic field.  For transient phenomena occurring during the peak of the magnetic field on the order of ps or ns after excitation by a fs pulse, the magnetic field variation is negligible.

Figure~\ref{ill:OpticalLayout} illustrates the optical setup for both time-resolved PL using the optical Kerr gate method\cite{Ho84,ArzhantsevMarconcelli05AS} using the amplified Ti:sapphire laser.  The PL is collected and collimated with an off-axis parabolic mirror after passing through the window on the magnet cryostat side.  This mirror can be adjusted to maximize the PL collected from either directly behind the excitation spot (center-collection) or from a $\mu$-prism at the edge of the sample to redirect the in-plane emission (edge-collection). This PL is focused onto a Kerr medium, toluene, with a second off-axis parabolic mirror.  After the Kerr medium, the PL is collimated again and focused into an optical fiber and measured with a silicon CCD attached to a grating spectrometer located over 2~meters away from the magnet.  Two crossed polarizers are located in the path of the PL, one before the Kerr medium and one after the Kerr medium, to block any PL that travels through the collection path without a gate pulse incident on the Kerr medium.  A gate pulse is focused onto the Kerr medium overlapping with the PL in space with polarization rotated 45$^\circ$ with respect to the two crossed polarizers.  A long wavelength pass filter is used to prevent scattered light of the gate pulse from entering the collection fiber.  To capture the emission dynamics, the gate pulse is incrementally delayed with respect to the PL, using a linear stage in the path of the gate beam.  During the Kerr medium's response to the gate pulse, the Kerr medium acts as a waveplate rotating the polarization of the PL, allowing some of the PL to pass the second polarizer into the collection.  In this way, the time-resolved PL can be mapped out as a function of wavelength and time.

\subsubsection{Time-integrated photoluminescence and absorption spectroscopy}

The time-integrated PL is collected with the same optical setup, except that the crossed polarizers and the Kerr medium are removed and the gate beam is blocked.  Because the PL emission from our sample occurs on a time scale of ns or less, the magnetic field is essentially held constant during the excitation and emission processes.  Under this scheme, the time-integrated PL is collected upon a single laser shot and single magnetic field shot.  
In addition to the strong excitation measurements, we use a laser diode (World Star Tech, TECiRL-15G-780) centered at 780~nm for weak excitation PL measurements.  The optical layout for the weak excitation PL measurements is the same as the time-integrated PL measurements described above, except a laser diode is used instead of the amplified Ti:sapphire laser.  The laser diode is modulated to be turned on for 400~$\mu$s at the peak of the magnetic field.  The magnetic field variation at the peak of the field for this time period is $\sim$4 parts in 300, less than 2~\% [see Fig.~\ref{ill:Sync}(b)].

For transmission measurements, we use an LED (Thorlabs, Inc., LED880L-50) centered at 880~nm with emission from 800~nm to 970~nm; the optical layout for the transmission measurements is not shown.  We take two transmission measurements while the sample is located in the magnet:  one at a high magnetic field, $T(B)_{\rm In~Mag}$, and the other at $B=0$, $T(B=0)_{\rm In~Mag}$.  Then, we take two more measurements using the helium flow cryostat without connecting the magnet cryostat.  One transmission measurement with the sample placed over an aperture where the sample is at the same temperature as it was in the magnet, $T(B=0)$, and another measurement of the incident light without the sample in the light path, $T_{\rm Inc}$.  The reference spectrum, $T_{Ref}$, is calculated from $T_{\rm Inc}$ by assuming a constant wavelength dependence for the reflection loss for GaAs at 12.5~K and the sapphire plate that the sample was mounted on.  The transmittance is calculated by $T(B)=\frac{T(B=0)}{T_{\rm Ref}} \frac{T(B)_{\rm In~Mag}}{T(B=0)_{\rm In~Mag}}$. Finally, we calculate the absorbance, $A(B)$, as $A(B)=-\log(T(B))$.

\section{Experimental results and discussion}
\label{sec:ExpRes}

\begin{figure}
\includegraphics[scale=0.62]{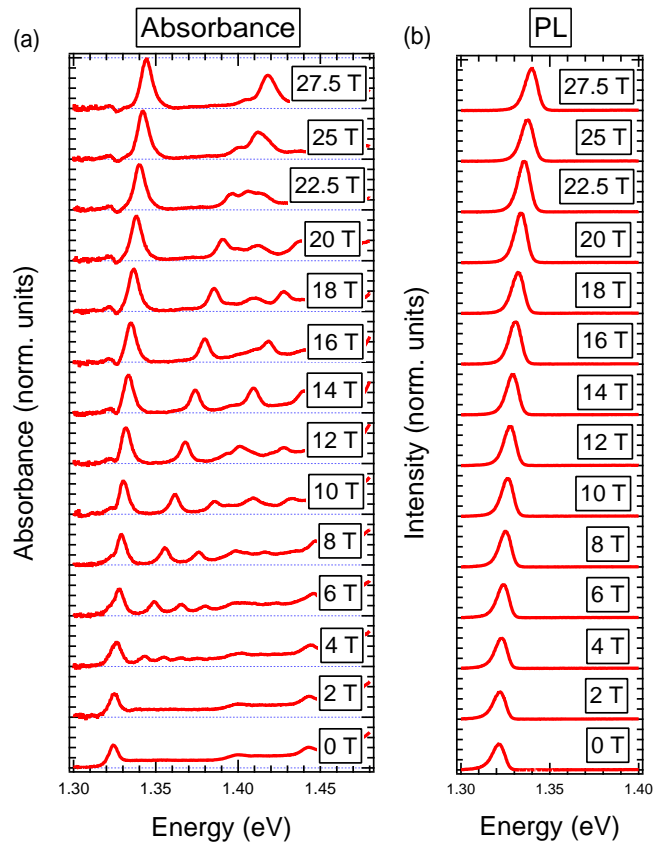}
\caption{(a)~Absorbance spectra taken using a light emitting diode, and (b)~weak excitation photoluminescence spectra using a laser diode, for undoped InGaAs quantum wells as a function of magnetic field at 12.5~K.}
\label{ill:LowDensity}
\end{figure}


Figure~\ref{ill:LowDensity} displays the absorbance and weak excitation PL spectra as a function of magnetic field using the LED and laser diode, respectively.  At 0~T, the absorbance shows a stair-step profile with excitonic peaks located near the subband edges, typical of a quasi-2D quantum well sample.\cite{Dingle75}  The subband edges can be identified as the E$_{1}$H$_{1}$ located at 1.325~eV, the E$_{1}$L$_{1}$ located at 1.4~eV, and the E$_{2}$H$_{2}$ located at 1.442~eV.  The large separation, 75~meV, between the E$_{1}$H$_{1}$ and E$_{1}$L$_{1}$ edges is due to strain\cite{MenendezetAl87PRB} between the In$_{0.2}$Ga$_{0.8}$As and the GaAs layers.  With increasing magnetic field, Landau quantization causes the quasi-2D density of states to evolve into a series of delta functions reminiscent of a quasi-0D density of states as the electron motion becomes fully quantized with the application of a high magnetic field perpendicular to the quantum wells.  The weak excitation PL spectra show a single peak of emission near the band edge that increases in strength and shifts to higher energy with increasing magnetic field (diamagnetic shift\cite{ShinadaTanaka70JPSJ,MacDonaldRitchie86PRB}).

\begin{figure}
\includegraphics[scale=0.62]{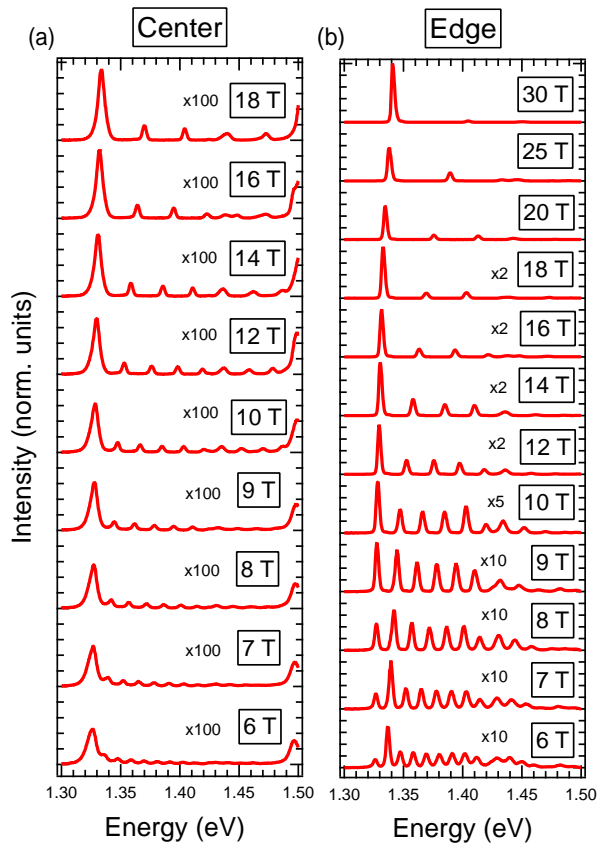}
\centering \caption{Time-integrated photoluminescence spectra upon intense excitation using an amplified Ti:sapphire laser for both (a)~center- and (b)~edge-emission with 5~$\mu$J excitation pulse energy at 13~K.}
\label{ill:TIPLSpectra}
\end{figure}

Figure~\ref{ill:TIPLSpectra} displays the results of the time-integrated PL spectra for both center- and edge-collected emission taken at $\sim$13~K with 5~$\mu$J excitation pulse energy using the amplified Ti:sapphire laser.  At each magnetic field and for each emission direction, four single-shot measurements were taken.  The displayed results are in each case the average spectrum of the four measurements.  The spectra are normalized with respect to the collected light by the CCD for the edge-collected emission at 30~T and no geometrical considerations were used regarding the collection direction, center or edge.  The peak located at $\sim$1.50~eV in the center-collected data is emission from the GaAs barriers and/or substrate.  In both sets of spectra, we see multiple Landau level (LL) peaks, which increase in separation with increasing magnetic field.  In the center collected spectra, the emission strength increases steadily for all of the LL transitions arising out of the E$_{1}$H$_{1}$ transition with increasing magnetic field. However, in the edge-collected spectra, we see a dramatic increase in emission strength from 6 to 30~T for the 00~LL.  The dramatic increase in intensity between the center- and edge-emission illustrate the fact that we are observing stimulated emission, or superfluorescence,\cite{JhoetAl06PRL,JhoetAl10PRB,NoeetAl12NP,NoeetAl13FP,KimetAl13PRB} from a dense electron-hole plasma for the in-plane direction.  The intensity of the edge-collected emission would be less than the center-collected emission if both were typical spontaneous emission because of the geometry of the collection.

\begin{figure}
\includegraphics[scale=0.75]{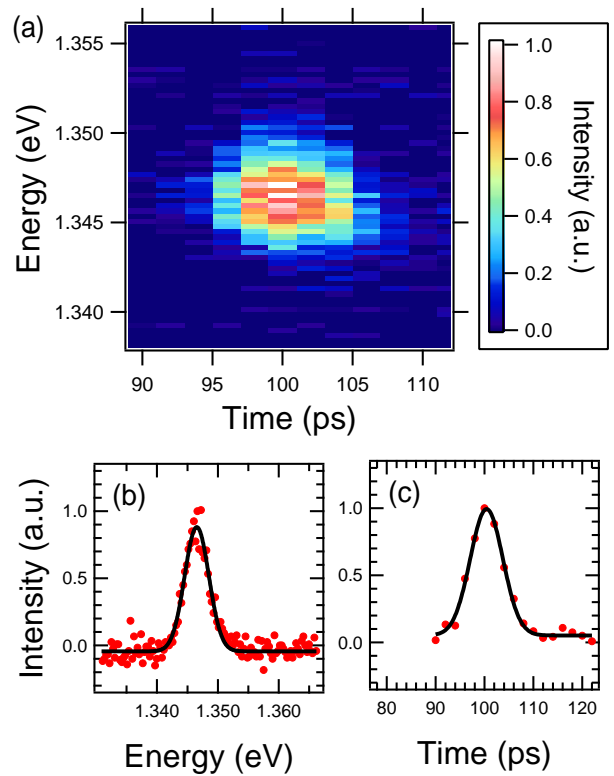}
\centering \caption{Superfluorescent burst of radiation from a highly excited InGaAs quantum well sample at 10~T measured using the mini-coil magnet.  A zoomed version, (a), shows the details of the data quality, and we show spectral, (b), and temporal, (c), slices of the data.}
\label{ill:TRPLat10Tesla}
\end{figure}

Figure~\ref{ill:TRPLat10Tesla} displays the result of the time-resolved PL map at 10~T and at 19~K when measuring the edge-collected emission.  After launching a series of magnet pulses, we partially created a map showing a burst of emission from the 11~LL energy.  Taking vertical and horizontal slices at the peak of the burst, we can determine the pulse duration to be $\sim$10~ps and spectral width to be $\sim$5~meV.  In our previous measurements at the National High Magnetic Field Laboratory, our temporal resolution for time-resolved PL measurements was limited to 20~ps due to dispersion in the graded-index fiber that was used for collection.\cite{NoeetAl12NP,NoeetAl13FP,KimetAl13PRB}  Here, we are able to place an upper limit for the SF pulse as the temporal resolution is limited by the Kerr medium, which, for toluene, is 1~ps.\cite{ChenetAl12APL}

\label{sec:TRPL}

\section{Conclusion}
\label{sec:Concl}
We have developed a unique mini-coil magnet system for nonlinear and ultrafast optical spectroscopy studies of materials.  Using this system, we have extended our most recent studies on superfluorescence from a high-density electron-hole plasma in semiconductor quantum wells\cite{NoeetAl12NP,NoeetAl13FP,KimetAl13PRB} to higher magnetic field strengths and with better temporal resolution for the time-resolved PL results by developing this system.

From a more general perspective, this unique magneto-optical spectroscopy system will open doors to many new types of experiments in condensed matter systems at high magnetic fields. Depending on the data acquisition speed, experiments that require the magnetic field dependence can be swept from 0-30~T within a single magnet pulse, and then repeated to improve the signal-to-noise ratio by averaging. The optical access via interchangeable windows allows us to introduce a variety of wavelengths, and, importantly, the application most suited for this magnet will be time-domain terahertz spectroscopy~\cite{IkebeetAl10PRL,ShimanoetAl13NC,BordacsetAl12NP,TakahashietAl11NP,WangetAl07OL,WangetAl10NP,WangetAl10OE,ArikawaetAl11PRB,ArikawaetAl12OE} because of the compact design and direct optical access to the sample. Furthermore, the direct optical access allows polarization-sensitive measurements\cite{ArikawaetAl12OE,ArikawaetAl13JIMT} without the complications that arise with optical fibers. Finally, the mini-coil design can be reproduced by other researchers around the world and incorporated into setups that already use expensive ultrafast laser systems or other sophisticated optical systems, greatly expanding the availability of high magnetic fields for condensed matter and materials research.

%
%

%

\begin{acknowledgments}
We acknowledge support from the National Science Foundation (through Grant No.~DMR-1310138), the Department of Energy (through Grant No.~DE-FG02-06ER46308), and the Robert A.~Welch Foundation (through Grant No.~C-1509)
\end{acknowledgments}


%

\end{document}